\documentclass[a4paper,12pt]{article}%
\usepackage{graphicx}
\usepackage{amssymb}
\usepackage{amsfonts}
\usepackage{amsmath}
\usepackage[font={small},labelsep=period]{caption}
\usepackage{cite}%
\providecommand{\U}[1]{\protect\rule{.1in}{.1in}}

\begin{document}

\author{M. S. Benilov$^{1}$ and D. M. Thomas$^{2}$\\$^{1}$Departamento de F\'{\i}sica, CCCEE, Universidade da Madeira, \\Largo do Munic\'{\i}pio, 9000 Funchal, Portugal\smallskip\\$^{2}$Blackett Laboratory, Imperial College London, Prince Consort Road,\\London, SW7 2BW, UK}
\title{Asymptotic theory of double layer and shielding of electric field at the edge
of illuminated plasma}
\date{}
\maketitle

\begin{abstract}
The method of matched asymptotic expansions is applied to the problem of a
collisionless plasma generated by UV illumination localized in a central part
of the plasma in the limiting case of small Debye length $\lambda_{D}$. A
second-approximation asymptotic solution is found for the double layer
positioned at the boundary of the illuminated region and for the
un-illuminated plasma for the plane geometry. Numerical calculations for
different values of $\lambda_{D}$ are reported and found to confirm the
asymptotic results. The net integral space charge of the double layer is
asymptotically small, although in the plane geometry it is just sufficient to
shield the ambipolar electric field existing in the illuminated region and
thus to prevent it from penetrating into the un-illuminated region. The double
layer has the same mathematical nature as the intermediate transition layer
separating an active plasma and a collisionless sheath, and the underlying
physics is also the same. In essence, the two layers represent the same
physical object: a transonic layer.

\end{abstract}

\section{Introduction}

In the first part of this work \cite{2013i} a collisionless plasma, generated
by UV illumination localized in a central part of the plasma, was analyzed.
The ions were assumed to be cold and the fluid description was used. Both
plane and cylindrical geometries were treated. An approximate analytical
solution was found under the approximation of quasi-neutrality and the exact
solution was computed numerically for one value of the Debye length
$\lambda_{D}$ for each geometry, this value being much smaller than widths of
both illuminated and un-illuminated regions. It was found that the ions
generated in the illuminated region are accelerated up to approximately the
Bohm speed inside the illuminated region. In plane geometry, the ions flow
across the un-illuminated region towards the near-wall positive space-charge
sheath with a speed which is virtually constant and slightly exceeds the Bohm
speed. In cylindrical geometry, the ions continue to be accelerated in the
un-illuminated region and enter the near-wall space-charge sheath with a speed
significantly exceeding the Bohm speed. In both geometries, a double layer
forms where the illuminated and un-illuminated regions meet.

A very unusual, if not unique, feature that this simple system reveals in
plane geometry is the coexistence of two quasi-neutral plasmas of the same
size with the ambipolar electric field being confined in one of them (the
illuminated plasma), while the other (the un-illuminated plasma) is to high
accuracy electric field-free and uniform. The latter is particularly
surprising since in all known models a near-wall space-charge sheath is
bordered by a nonuniform quasi-neutral presheath where the ions going to the
sheath are accelerated and the voltage drop is of the order of the electron
temperature measured in volts. (We set aside cases where the ions are produced
on the surface, as in Q-machines or experiments with heated cavities
\cite{Phelps1976}, or inside the sheath, as in near-cathode layers of
discharges burning in cathode vapour \cite{2010f}.) Moreover, the difference
between illuminated and un-illuminated regions in terms of ion momentum is in
the presence or absence of ionization friction force, and the fact that the
ion fluid is accelerated in the illuminated plasma, where the ionization
friction force is present, and is not accelerated in the un-illuminated
plasma, where the friction force is absent, is somehow counterintuitive.

The above feature is extremely interesting, also from the methodological point
of view; note that the classical Bohm sheath solution \cite{Bohm1949} is
sufficient to describe both the sheath and the adjacent (un-illuminated)
plasma in such situation. A key to this feature is the double layer, which
shields the ambipolar electric field induced in the illuminated region and
prevents it from penetrating the un-illuminated region.

The quasi-neutral analytical solution \cite{2013i}, while clearly being
useful, does not describe the double layer, hence a more sophisticated
treatment is needed in order to fully understand this feature and the
underlying physics. It is clear that this feature originates in the smallness
of $\lambda_{D}$, therefore the relevant procedure is to find an asymptotic
solution to the considered problem in the limiting case of small $\lambda_{D}%
$. The technique of choice to this end is the method of matched asymptotic
expansions (e.g.,
\cite{van-Dyke1964,Cole1968,Nayfeh1973,Nayfeh1981,Kevorkian1981,Nayfeh1985}),
which is a standard tool in problems with singular perturbations. Note that
this method has been successfully used in the theory of plasma-wall
transitions in collisionless plasmas; e.g., reviews
\cite{Riemann1991,Franklin2003c,Allen2009,2009a,Riemann2009a}. In particular,
relevant in the present context are works \cite{Lam1965,Franklin1970}, where a
transition layer separating active plasma and a collisionless sheath was
introduced; \cite{2012b}, where a more adequate and simpler mathematical
description of this layer was suggested; and \cite{1999d}, where the plasma
column in electronegative gases was studied including in the exceptional case
where the column comprises an inner plasma, a double layer, an outer plasma,
and a near-wall sheath.

An approximate analytical solution in the limiting case of small $\lambda_{D}$
found by means of the method of matched asymptotic expansions is reported in
this paper, which is thus complementary to \cite{2013i}. Also reported are
results of numerical calculations for different values of $\lambda_{D}$.
Questions to be answered include: what is the physics of the double layer; why
the ion speed in the un-illuminated plasma deviates from the Bohm speed; how
this deviation can be estimated.

The outline of the paper is as follows. Equations and boundary conditions are
briefly described in Sec.\ \ref{Equations}. An asymptotic solution for plane
geometry is given and compared with numerical results in
Sec.\ \ref{Asymptotic solution}. An asymptotic solution for cylindrical
geometry is outlined in Sec.\ \ref{Cylindrical geometry}. A concluding
discussion is given in Sec.\ \ref{Discussion}. Mathematical details are placed
in two Appendices in order not to overload the text.

\section{Equations and boundary conditions}

\label{Equations}

We refer to \cite{2013i} for the description of the physical situation. In
brief, we consider a plane or cylindrical plasma produced by UV radiation.
Governing equations are written in the fluid approximation and are well-known
and the same as in \cite{2013i}; they include the ion conservation equation,
the ion momentum equation written without account of collisions, the
equilibrium equation for the electrons, and the Poisson equation:%
\begin{equation}
\frac{1}{x^{\beta}}\frac{d}{dx}\left(  x^{\beta}n_{i}v_{i}\right)
=G,\label{1}%
\end{equation}%
\begin{equation}
m_{i}n_{i}v_{i}\frac{dv_{i}}{dx}=en_{i}E-Gm_{i}v_{i}.\label{2a}%
\end{equation}%
\begin{equation}
\frac{d}{dx}\left(  n_{e}kT_{e}\right)  +en_{e}E=0,\label{3}%
\end{equation}%
\begin{equation}
\frac{\varepsilon_{0}}{x^{\beta}}\frac{d\left(  x^{\beta}E\right)  }%
{dx}=e\left(  n_{i}-n_{e}\right)  ,\label{4}%
\end{equation}
where $\beta=0$ and $x$ is the distance from the plane of symmetry for plane
geometry, $\beta=1$ and $x$ is the distance from the axis for cylindrical
geometry, $G$ is the ionization rate, and other designations are the usual ones.

It is convenient for the purposes of this work to replace the ion momentum
equation Eq.\ (\ref{2a}) by an equivalent equation which can be derived
following \cite{2000a,2012b} and reads%
\begin{equation}
\left(  \frac{v_{i}}{c_{s}^{2}}-\frac{1}{v_{i}}\right)  \frac{dv_{i}}%
{dx}=\frac{\varepsilon_{0}}{en_{i}}\left\{  \frac{d}{dx}\left[  \frac
{1}{x^{\beta}}\frac{d\left(  x^{\beta}E\right)  }{dx}\right]  +\frac
{eE}{kT_{e}x^{\beta}}\frac{d\left(  x^{\beta}E\right)  }{dx}\right\}
-\frac{G}{c_{s}n_{i}}\left(  \frac{v_{i}}{c_{s}}+\frac{c_{s}}{v_{i}}\right)
+\frac{\beta}{x}.\label{6}%
\end{equation}
Here $c_{s}=\sqrt{kT_{e}/m_{i}}$ is the Bohm speed. Note that for $\beta=0$
this equation coincides with the corresponding equation from \cite{2012b}
except that Eq.\ (\ref{6}) does not account for ion-atom collisions.

We consider a situation where the illumination is localized in a central part
of the plasma and is uniform there, so $G=G_{0}$ for $0\leq x<\Delta$ and
$G=0$ for $\Delta<x\leq L$, where $G_{0}$ is a (positive) constant, $\Delta$
is the halfwidth of irradiated region and $L$ is the halfwidth of the system
in the plane case, and $\Delta$ is the radius of the irradiated region and $L$
is the discharge tube radius in the cylindrical case. Note that $n_{i}$,
$n_{e}$, $v_{i}$, and electrostatic potential $\varphi$ are, of course,
continuous at $x=\Delta$; the derivatives $d\varphi/dx$ and $d^{2}%
\varphi/dx^{2}$ are also continuous, which follows from Eq.\ (\ref{4}); the
derivative $dn_{e}/dx$ is continuous as well, which follows from
Eq.\ (\ref{3}); the derivatives $dv_{i}/dx$, $dn_{i}/dx$, and $d^{3}%
\varphi/dx^{3}$ are discontinuous, which follows from, respectively,
Eqs.\ (\ref{2a}), (\ref{1}), and (\ref{4}).

Boundary conditions are also well-known and the same as those in
\cite{2013i}:
\begin{equation}
x=0:\quad\quad\frac{dn_{i}}{dx}=0,\quad\quad v_{i}=0,\quad\quad\varphi
=0,\quad\quad E=0;
\end{equation}%
\begin{equation}
x=L:\quad\quad n_{i}v_{i}=\frac{1}{4}n_{e}\left(  \frac{8kT_{e}}{\pi m_{e}%
}\right)  ^{1/2}.\label{8}%
\end{equation}

Eq.\ (\ref{1}) can be integrated directly and the result may be written in
terms of the Heaviside step function $H\!\left(  x\right)  $ ($H\!\left(
x\right)  =0$ for $x<0$; $H\!\left(  x\right)  =1$ for $x>0$):%
\begin{equation}
n_{i}v_{i}=\frac{G_{0}x}{1+\beta}+\frac{G_{0}}{1+\beta}\left(  \frac
{\Delta^{1+\beta}}{x^{\beta}}-x\right)  H\!\left(  x-\Delta\right)  .
\end{equation}

Introduce dimensionless variables
\begin{equation}
\xi=\frac{x}{\Delta},\quad\quad N_{i}=\frac{n_{i}}{n_{0}},\quad\quad
N_{e}=\frac{n_{e}}{n_{0}},\quad\quad V=\frac{v_{i}}{c_{s}},\quad\quad
\Phi=\frac{e\varphi}{kT_{e}},\label{5.1}%
\end{equation}
where $n_{0}=G_{0}\Delta/\left(  1+\beta\right)  c_{s}$ is a characteristic
density of the charged particles. (Note that this normalization differs from
the one used in \cite{2013i} in that it employs normalization factors
involving only control parameters.) The governing equations assume the form%
\begin{equation}
N_{i}V=\xi+\left(  \frac{1}{\xi^{\beta}}-\xi\right)  \,H\!\left(
\xi-1\right)  ,\label{5.2}%
\end{equation}%
\begin{align}
\frac{V^{2}-1}{V}\frac{dV}{d\xi}  & =\frac{\varepsilon}{N_{i}}\left\{
-\frac{d}{d\xi}\left[  \frac{1}{\xi^{\beta}}\frac{d}{d\xi}\left(  \xi^{\beta
}\frac{d\Phi}{d\xi}\right)  \right]  +\frac{d\Phi}{d\xi}\frac{1}{\xi^{\beta}%
}\frac{d}{d\xi}\left(  \xi^{\beta}\frac{d\Phi}{d\xi}\right)  \right\}
\label{ion_mom}\\
& -\frac{1+\beta}{\xi}\left(  V^{2}+1\right)  H\!\left(  1-\xi\right)
+\frac{\beta}{\xi},\nonumber
\end{align}%
\begin{equation}
\frac{dN_{e}}{d\xi}-N_{e}\frac{d\Phi}{d\xi}=0,
\end{equation}%
\begin{equation}
\frac{\varepsilon}{\xi^{\beta}}\frac{d}{d\xi}\left(  \xi^{\beta}\frac{d\Phi
}{d\xi}\right)  =N_{e}-N_{i},\label{5.5}%
\end{equation}
where $\lambda_{D}=\left(  \varepsilon_{0}kT_{e}/e^{2}n_{0}\right)  ^{1/2}$
and $\varepsilon=\left(  \lambda_{D}/\Delta\right)  ^{2}$.

The next section, Sec.\ \ref{Asymptotic solution}, is concerned with plane
geometry, $\beta=0$. The asymptotic solution for cylindrical geometry,
$\beta=1$, is outlined in Sec.\ \ref{Cylindrical geometry}.

\section{Asymptotic solution: plane geometry}

\label{Asymptotic solution}

\subsection{Asymptotic structure of the solution}

\label{The approach}

Asymptotic zones that need to be considered may be conveniently illustrated by
invoking results of numerical calculations. The calculations have been
performed in the same way as in \cite{2013i} and an example is depicted by
solid lines in Fig.\ 1. One can clearly see four regions with apparently
different physics: the illuminated plasma, $0\leq\xi<1$; the un-illuminated
plasma, $1<\xi<2$; a thin intermediate (double) layer positioned in the
vicinity of the point $\xi=1$ and separating the two plasmas; and a thin
space-charge sheath positioned at the wall, i.e., at $\xi$ close to $2$.

In the course of application of the method of matched asymptotic expansions
the same four regions appear as asymptotic zones described by different
asymptotic expansions. The illuminated and un-illuminated plasmas are
described by straightforward expansions, i.e., the first term of each of these
expansions is of the order unity and governed by equations which are obtained
from Eqs.\ (\ref{5.2})-(\ref{5.5}) by setting $\varepsilon=0$. The latter is,
of course, consistent with the plasmas in both the illuminated and
un-illuminated regions being quasi-neutral.

The double layer is described by the same asymptotic expansion as the one
which describes the transition layer separating active plasma and a
collisionless sheath \cite{Franklin1970}. The first term of the expansions of
each of the quantities $V$, $N_{i}$, $N_{e}$, and $\Phi$ is constant.
Therefore, the double layer appears only in the second approximation and may
be ignored in the first one.

The near-wall space-charge sheath is to the first approximation the well-known
Bohm space-charge sheath. Its asymptotic description in the present problem is
similar to the description given in \cite{Bohm1949} and is skipped for brevity.

\subsection{The first approximation}

\label{The first approximation}

The region of illuminated plasma, $0\leq\xi<1$, is described by the
straightforward expansion%
\begin{equation}
V\left(  \xi,\varepsilon\right)  =V_{1}\left(  \xi\right)  +\alpha_{1}%
V_{2}\left(  \xi\right)  +\dots,\label{10}%
\end{equation}%
\begin{equation}
N_{i}\left(  \xi,\varepsilon\right)  =\frac{\xi}{V_{1}\left(  \xi\right)
}\left[  1-\alpha_{1}\frac{V_{2}\left(  \xi\right)  }{V_{1}\left(  \xi\right)
}\right]  +\dots,
\end{equation}%
\begin{equation}
N_{e}\left(  \xi,\varepsilon\right)  =N_{e1}\left(  \xi\right)  +\alpha
_{1}N_{e2}\left(  \xi\right)  +\dots,
\end{equation}%
\begin{equation}
\Phi\left(  \xi,\varepsilon\right)  =\Phi_{1}\left(  \xi\right)  +\alpha
_{1}\Phi_{2}\left(  \xi\right)  +\dots,\label{13}%
\end{equation}
where $\alpha_{1}=\alpha_{1}\left(  \varepsilon\right)  $ is a small parameter
which is to be found in the course of analysis as a part of solution.

Substituting expansion (\ref{10})-(\ref{13}) into Eqs.\ (\ref{ion_mom}%
)-(\ref{5.5}), expanding and retaining the leading terms, one obtains
equations%
\begin{equation}
\frac{V_{1}^{2}-1}{V_{1}}\frac{dV_{1}}{d\xi}=-\frac{1}{\xi}\left(  V_{1}%
^{2}+1\right)  \,,
\end{equation}%
\begin{equation}
\frac{dN_{e_{1}}}{d\xi}-N_{e1}\frac{d\Phi_{1}}{d\xi}=0,
\end{equation}%
\begin{equation}
N_{e1}-\frac{\xi}{V_{1}}=0.
\end{equation}

A solution subject to boundary condition $\Phi_{1}\left(  0\right)  =0$ reads%
\begin{equation}
V_{1}=\frac{C_{1}\xi}{1+\sqrt{1-\left(  C_{1}\xi\right)  ^{2}}},\;\;\;\;N_{e1}%
=\frac{1+\sqrt{1-\left(  C_{1}\xi\right)  ^{2}}}{C_{1}},\;\;\;\Phi_{1}%
=\ln\frac{1+\sqrt{1-\left(  C_{1}\xi\right)  ^{2}}}{2},
\end{equation}
where $C_{1}$ is an integration constant, $0<C_{1}\leq1$. Note that
$V_{1}\left(  1\right)  $ (i.e., the value of the function $V_{1}\left(
\xi\right)  $ for $\xi=1$) increases with increasing $C_{1}$ and its maximum
value is attained at $C_{1}=1$ and equals $1$.

The region of un-illuminated plasma, $1<\xi<S=L/\Delta$, is described by the
straightforward expansion
\begin{equation}
V\left(  \xi,\varepsilon\right)  =V_{3}\left(  \xi\right)  +\alpha_{2}%
V_{4}\left(  \xi\right)  +\dots,\label{22}%
\end{equation}%
\begin{equation}
N_{i}\left(  \xi,\varepsilon\right)  =\frac{1}{V_{3}\left(  \xi\right)
}\left[  1-\alpha_{2}\frac{V_{4}\left(  \xi\right)  }{V_{3}\left(  \xi\right)
}\right]  +\dots,
\end{equation}%
\begin{equation}
N_{e}\left(  \xi,\varepsilon\right)  =N_{e3}\left(  \xi\right)  +\alpha
_{2}N_{e4}\left(  \xi\right)  +\dots,
\end{equation}%
\begin{equation}
\Phi\left(  \xi,\varepsilon\right)  =\Phi_{3}\left(  \xi\right)  +\alpha
_{2}\Phi_{4}\left(  \xi\right)  +\dots,\label{25}%
\end{equation}
where $\alpha_{2}=\alpha_{2}\left(  \varepsilon\right)  $ is a small parameter
which is to be found as a part of solution. The leading terms of this
expansion are governed by the equations
\begin{equation}
\left(  V_{3}-\frac{1}{V_{3}}\right)  \frac{dV_{3}}{d\xi}=0,\label{26}%
\end{equation}%
\begin{equation}
\frac{dN_{e3}}{d\xi}-N_{e3}\frac{d\Phi_{3}}{d\xi}=0,
\end{equation}%
\begin{equation}
N_{e3}-\frac{1}{V_{3}\left(  \xi\right)  }=0.\label{28}%
\end{equation}

Expansion (\ref{22})-(\ref{25}) is to be matched with a double layer expansion
in the vicinity of the point $\xi=1$ and with a sheath expansion in the
vicinity of the point $\xi=S$. However, the double layer needs to be taken
into account only in the second and higher approximations, so the first terms
of the expansions describing the illuminated and un-illuminated plasmas should
be patched directly, meaning that $V_{1}\left(  1\right)  =V_{3}\left(
1\right)  $, $N_{e1}\left(  1\right)  =N_{e3}\left(  1\right)  $, $\Phi
_{1}\left(  1\right)  =\Phi_{3}\left(  1\right)  $. Solutions to
Eqs.\ (\ref{26})-(\ref{28}) subject to these boundary conditions are trivial:
$V_{3}=V_{1}\left(  1\right)  $, $N_{e3}=N_{e1}\left(  1\right)  $, $\Phi
_{3}=\Phi_{1}\left(  1\right)  $.

In order that a matching with the sheath expansion be possible, $V_{3}$ must
satisfy the Bohm criterion, i.e., it should be $V_{3}\geq1$. Since
$V_{1}\left(  1\right)  \leq1$ as discussed above, it follows that
$V_{1}\left(  1\right)  =V_{3}=1$ and $C_{1}=1$. It follows also that
$N_{e3}=1$, $\Phi_{3}=-\ln2$.

The first-approximation solution is complete now. In \cite{2013i}, the same
solution was obtained by patching solutions for the illuminated and
un-illuminated plasmas obtained with the use of the condition of
quasi-neutrality and the Bohm criterion.

This solution is depicted by the dotted lines in Fig.\ 1. As expected, it
provides a reasonable approximation of the ion speed. However, in the double
layer this approximation is not smooth, and the approximation of the electric
field is not suitable altogether.

Poor approximation of derivatives of approximate solutions is a deficiency
inherent to patching. In the method of matched asymptotic expansions, this
deficiency can be removed by going to the second approximation.

\subsection{The second approximation}

\label{The second approximation}

It is natural to seek an asymptotic expansion describing the double layer in
the same form as that of the expansion which describes the transition layer
separating active plasma and a collisionless sheath \cite{Franklin1970}:%
\begin{equation}
V\left(  \xi,\varepsilon\right)  =1+\varepsilon^{1/5}V_{5}\left(  \xi
_{5}\right)  +\dots,\label{30}%
\end{equation}%
\begin{equation}
N_{i}\left(  \xi,\varepsilon\right)  =1-\varepsilon^{1/5}V_{5}\left(  \xi
_{5}\right)  +\dots,\label{31}%
\end{equation}%
\begin{equation}
N_{e}\left(  \xi,\varepsilon\right)  =1+\varepsilon^{1/5}N_{e5}\left(  \xi
_{5}\right)  +\dots,\label{32}%
\end{equation}%
\begin{equation}
\Phi\left(  \xi,\varepsilon\right)  =-\ln2+\varepsilon^{1/5}\Phi_{5}\left(
\xi_{5}\right)  +\dots,\label{33}%
\end{equation}
where $\xi_{5}=\left(  \xi-1\right)  /\varepsilon^{2/5}$. The form of this
expansion was derived in \cite{Franklin1970} with the use of considerations
stemming from matching and in \cite{2012b} by means of estimates of different
terms of equation (\ref{6}). A convenient illustration of this expansion is
provided by Fig.\ 1: the expansion implies that variations of the quantities
$V$, $N_{i}$, $N_{e}$, and $\Phi$ in the double layer are of the order of
$\varepsilon^{1/5}$ and the electric field is of the order of $\varepsilon
^{-1/5}$, and this is consistent with Fig.\ 1.

Substituting this expansion into Eqs.\ (\ref{ion_mom})-(\ref{5.5}), expanding
and retaining the leading terms, one obtains equations%
\begin{equation}
2V_{5}\frac{dV_{5}}{d\xi_{5}}=-\frac{d^{3}\Phi_{5}}{d\xi_{5}^{3}}-2H\!\left(
-\xi_{5}\right)  ,\label{34}%
\end{equation}%
\begin{equation}
\frac{dN_{e5}}{d\xi_{5}}-\frac{d\Phi_{5}}{d\xi_{5}}=0,\label{35}%
\end{equation}%
\begin{equation}
N_{e5}+V_{5}=0.\label{36}%
\end{equation}

As seen from Eq.\ (\ref{36}), the plasma in the double layer is quasi-neutral
not only to the first approximation, but also to the second one. A consequence
is that the derivative $dV_{5}/d\xi_{5}$ is continuous at $\xi_{5}=0$. Coming
back to the discussion of Sec.\ \ref{Equations}, one can say that the
discontinuity of derivative $dv_{i}/dx$ at $x=\Delta$ occurs not in the first
approximation in $\varepsilon$ but rather in the second one. The latter is
consistent with the second term on the rhs of Eq.\ (\ref{2a}), which is
responsible for the discontinuity, being of the order of $\varepsilon^{1/5}$,
i.e., asymptotically small, with respect to the first term.

Eqs.\ (\ref{34}) and (\ref{35}) may be integrated term by term to give%
\begin{equation}
V_{5}^{2}+\frac{d^{2}\Phi_{5}}{d\xi_{5}^{2}}+2\xi_{5}H\!\left(  -\xi
_{5}\right)  =C_{2}^{2},\label{37}%
\end{equation}%
\begin{equation}
N_{e5}-\Phi_{5}=C_{3},\label{38}%
\end{equation}
where $C_{2}^{2}$ and $C_{3}$ are integration constants. Note that since every
term on the lhs of these equations is continuous at $\xi=0$, the constants
$C_{2}^{2}$ and $C_{3}$ do not switch their values between the regions $\xi>0$
and $\xi<0$.

The system of Eqs.\ (\ref{37}), (\ref{38}), (\ref{36}) may be reduced to a
single equation, for example, for the unknown $V_{5}$:%
\begin{equation}
V_{5}^{2}=\frac{d^{2}V_{5}}{d\xi_{5}^{2}}-2\xi_{5}H\!\left(  -\xi_{5}\right)
+C_{2}^{2}.\label{43}%
\end{equation}
A solution of this equation subject to relevant boundary conditions, which
follow from the van Dyke asymptotic matching principle \cite{van-Dyke1964}, is
found in Appendix \ref{Appendix} and depicted in Fig.\ 2 by the solid line for
$\xi_{5}\leq0$ and dash-dotted line for $\xi_{5}\geq0$. (Here $X=\xi
_{5}-0.2254$.) The constant $C_{2}$ is found to be equal to $0.6714$. Also
shown in Fig.\ 2 is the function $V_{5}\left(  X\right)  $ referring to the
transition layer separating active plasma and a collisionless sheath
\cite{Franklin1970}. The two functions coincide in the range $X\leq-0.2254$
and differ for bigger $X$.

We will need for subsequent asymptotic matching the two-term asymptotic
expansion of the functions $V_{5}\left(  \xi_{5}\right)  $ and $N_{5}\left(
\xi_{5}\right)  $ for $\xi\rightarrow-\infty$, the three-term asymptotic
expansion of the function $\Phi_{5}\left(  \xi_{5}\right)  $ for
$\xi\rightarrow-\infty$, and the one-term expansion of all these functions for
$\xi\rightarrow\infty$. These expansions may be readily found:
\begin{align}
\xi_{5}\rightarrow-\infty:\;\;\; &  V_{5}=-\sqrt{-2\xi_{5}}-\frac{C_{2}^{2}%
}{\sqrt{-8\xi_{5}}}+\dots,\;\;\;N_{e5}=\sqrt{-2\xi_{5}}+\frac{C_{2}^{2}}%
{\sqrt{-8\xi_{5}}}+\dots,\;\;\;\label{42}\\
&  \Phi_{5}=\sqrt{-2\xi_{5}}-C_{3}+\frac{C_{2}^{2}}{\sqrt{-8\xi_{5}}}%
+\dots;\nonumber
\end{align}%
\begin{equation}
\xi_{5}\rightarrow\infty:\;\;\;V_{5}\rightarrow C_{2},\;\;\;N_{e5}%
\rightarrow-C_{2},~~~~\Phi_{5}\rightarrow-C_{2}-C_{3}.\label{48}%
\end{equation}

Let us now consider the asymptotic matching of expansions (\ref{30}%
)-(\ref{33}) and (\ref{10})-(\ref{13}) accounting for two terms in each
expansion. Making use of Eq.\ (\ref{42}), one finds that the matching is
possible provided that $\alpha_{1}=\varepsilon^{2/5}$, $C_{3}=0$, and
\begin{equation}
V_{2}=-\frac{C_{2}^{2}}{\sqrt{8\left(  1-\xi\right)  }}+\dots,\;\;\;\;N_{e2}%
=\frac{C_{2}^{2}}{\sqrt{8\left(  1-\xi\right)  }}+\dots,\;\;\text{\ }\Phi
_{2}=\frac{C_{2}^{2}}{\sqrt{8\left(  1-\xi\right)  }}+\dots\label{54}%
\end{equation}
for $\xi\rightarrow1-0$. Eq.\ (\ref{54}) represents a set of boundary
conditions for differential equations governing functions $V_{2}$, $N_{e2}$,
and $\Phi_{2}$, which can be obtained by substituting expansion (\ref{10}%
)-(\ref{13}) into Eqs.\ (\ref{ion_mom})-(\ref{5.5}), expanding, and retaining
second-order terms. However, these functions represent a correction of the
order of $\alpha_{1}=\varepsilon^{2/5}$, which is higher than the order to
which the solution is known in the double layer ($\varepsilon^{1/5}$). For
this reason, we leave finding these functions beyond the scope of this work
and only note that the boundary conditions (\ref{54}) are compatible with the
differential equations.

Let us now consider the asymptotic matching of expansions (\ref{30}%
)-(\ref{33}) and (\ref{22})-(\ref{25}) accounting for two terms in each
expansion. Taking into account Eq.\ (\ref{48}), one finds that the matching is
possible provided that $\alpha_{2}=\varepsilon^{1/5}$ and
\begin{equation}
V_{4}\left(  1\right)  =C_{2},\;\;\;N_{e4}\left(  1\right)  =-C_{2}%
,\;\;\;\Phi_{4}\left(  1\right)  =-C_{2}.\label{50a}%
\end{equation}

Substituting expansions (\ref{22})-(\ref{25}) into Eqs.\ (\ref{ion_mom}%
)-(\ref{5.5}), expanding, and retaining second-order terms, one obtains
equations%
\begin{equation}
V_{4}\frac{dV_{4}}{d\xi}=0,\label{50}%
\end{equation}%
\begin{equation}
\frac{dN_{e4}}{d\xi}-\frac{d\Phi_{4}}{d\xi}=0,\;\;\;N_{e4}+V_{4}=0.\label{52}%
\end{equation}
Solution of Eqs.\ (\ref{50}) and (\ref{52}) subject to the boundary conditions
Eq.\ (\ref{50a}) is trivial: $V_{4}=C_{2}$, $N_{e4}=-C_{2}$, $\Phi_{4}=-C_{2}$.

The above results ensure a description of the illuminated plasma region to the
double layer to the un-illuminated region with the error of the order of
$\varepsilon^{2/5}$ (which is, presumably, the order of the third terms of the
expansions describing the double layer and the un-illuminated region). These
results allow one to construct a composite expansion
\cite{van-Dyke1964,Cole1968,Nayfeh1973,Nayfeh1981,Kevorkian1981,Nayfeh1985}
uniformly valid in all these regions with the error of the order of
$\varepsilon^{2/5}$; see Appendix \ref{Composite expansions}.

Comparison of the above asymptotic solution with results of numerical
calculations is shown in Figs.\ 3-5. While analyzing Figs.\ 3b-3d, it should
be kept in mind that $\Phi_{5}=-V_{5}$, which follows from Eqs.\ (\ref{36})
and (\ref{38}). Dotted lines in Figs.\ 3 and 4 represent the data given by
Eqs.\ (\ref{61}), (\ref{61a}), and (\ref{62a}). Values of the ion speed in the
un-illuminated region shown in Fig.\ 5 have been taken at points where
$dV/d\xi$ attains the minimum value except in the case $\varepsilon
=0.99\times10^{-5}$, where variation of the ion speed in the un-illuminated
region was below the usual floating-point precision.

One can see the deviation between the asymptotic and numerical results is
reasonably small and decreases with decreasing $\varepsilon$.

\section{Cylindrical geometry}

\label{Cylindrical geometry}

One needs to consider the same four asymptotic zones as in the case of plane
geometry. A composite first-approximation solution uniformly valid from the
illuminated plasma to the double layer to the un-illuminated plasma is
governed by Eq.\ (\ref{ion_mom}) with $\varepsilon=0$ and $\beta=1$:%
\begin{equation}
\frac{V^{2}-1}{V}\frac{dV}{d\xi}=-\frac{2}{\xi}\left(  V^{2}+1\right)
H\!\left(  1-\xi\right)  +\frac{1}{\xi}.\label{8a}%
\end{equation}
A continuous single-valued solution of this equation may be written in the
implicit form:%
\begin{equation}
\xi=\left\{
\begin{tabular}
[c]{lll}%
$\frac{3^{3/4}V}{\left(  2V^{2}+1\right)  ^{3/4}}$ & for & $V\leq1$\\
$\frac{1}{V}\exp\frac{V^{2}-1}{2}$ & for & $V\geq1$%
\end{tabular}
\ \ \right.  .\label{12}%
\end{equation}
One can see that $V\left(  1\right)  =1$, i.e., the sonic point is positioned
at the edge of the illuminated plasma, as in the planar case. The difference
is that the solution (\ref{12}) is obtained without invoking the Bohm
criterion; note that Eq.\ (\ref{8a}) admits continuous solutions with
$V\left(  1\right)  \neq1$, however these solutions are multi-valued. Another
difference is that $V>1$ for $\xi>1$, i.e., the plasma continues to be
accelerated in the un-illuminated plasma.

Most formulas of Sec.\ \ref{The second approximation} referring to the double
layer may be readily generalized in order to become applicable to both plane
and cylindrical geometries. In particular, this includes replacing the first
term on the rhs of Eq.\ (\ref{33}) with $-\left(  2+\beta\right)  \left(
2+2\beta\right)  ^{-1}\ln\left(  2+\beta\right)  $ and writing Eq.\ (\ref{43})
in the form%
\begin{equation}
V_{5}^{2}=\frac{d^{2}V_{5}}{d\xi_{5}^{2}}-2\left(  1+\beta\right)  \xi
_{5}H\!\left(  -\xi_{5}\right)  +\beta\xi_{5}+C_{2}^{2}.\label{12a}%
\end{equation}
A major difference between the functions $V_{5}\left(  \xi_{5}\right)  $ for
plane and cylindrical geometries is in their asymptotic behavior for $\xi
_{5}\rightarrow\infty$: for $\beta=1$ the latter is governed by the term
$\beta\xi_{5}$ on the rhs of Eq.\ (\ref{12a}) and reads $V_{5}\approx\sqrt
{\xi_{5}}$ , which is similar to the first relation in (\ref{42}) rather than
(\ref{48}). Note that this difference does not lead to significantly different
behavior of the normalized electric field and space-charge density,
$dV_{5}/d\xi_{5}$ and $dV_{5}^{2}/d\xi_{5}^{2}$: both tend to zero as $\xi
_{5}\rightarrow\infty$, although algebraically rather than exponentially as in
plane geometry.

Expansions (\ref{10})-(\ref{13}) and (\ref{22})-(\ref{25}), describing regions
of, respectively, illuminated and un-illuminated plasmas, remain applicable in
cylindrical geometry except that $\alpha_{2}$ becomes equal to $\varepsilon
^{2/5}$.

\section{Concluding discussion}

\label{Discussion}

The four zones with different physics revealed by numerical calculations and
shown in Fig.\ 1, that is, the illuminated plasma, the double layer, the
un-illuminated plasma, and the near-wall space-charge sheath appear in a
natural way in the course of application of the method of matched asymptotic
expansions. The first-order terms of asymptotic expansions of $V$, $N_{i}$,
and $\Phi$ describing the illuminated plasma, the double layer, and the
un-illuminated plasmas represent the same quasi-neutral solution that was
found in \cite{2013i} by patching solutions for the illuminated and
un-illuminated plasmas and (in plane geometry) invoking the Bohm criterion.
The second-order terms are of the order of $\varepsilon^{2/5}$ in the
illuminated plasma; $\varepsilon^{1/5}$ in the double layer; and
$\varepsilon^{1/5}$ in the un-illuminated plasma in plane geometry and
$\varepsilon^{2/5}$ in cylindrical geometry. It is interesting to note that
these orders explain why the quasi-neutral solution for $V$, $N_{i}$, and
$\Phi$ in Fig.\ 1 of this work and Figs. 2 and 3 of \cite{2013i} is more
accurate in the illuminated plasma than in the double layer and why its
accuracy in the un-illuminated plasma is comparable to that in the double
layer in plane geometry and in the illuminated plasma in cylindrical geometry.

The double layer separating the illuminated and un-illuminated plasmas is
quasi-neutral: the separation of charges here is of the order of
$\varepsilon^{2/5}$, i.e., asymptotically small, although much bigger than in
the illuminated and un-illuminated plasmas. The net integral space charge (per
unit cross section) of the double layer is asymptotically small, which can be
readily seen: in the first approximation it is proportional to $\left.
dV_{5}/d\xi_{5}\right\vert _{-\infty}^{\infty}$ and equals zero by virtue of
boundary conditions (\ref{42}) and (\ref{48}). However, in plane geometry the
net integral space charge of the double layer is just sufficient to shield the
ambipolar electric field existing in the illuminated region and thus to
prevent it from penetrating the un-illuminated region. The effect of the
ionization friction force [the second term on the rhs of Eq.\ (\ref{2a})] is
as well asymptotically small in the double layer, therefore the full energy of
an ion is conserved as reflected by the relationship $V_{5}+\Phi_{5}=0$.

The double layer has the same mathematical nature as the intermediate
transition layer separating an active plasma and a collisionless sheath, which
was introduced in \cite{Lam1965,Franklin1970} and revisited in, e.g.,
\cite{Riemann1997,Kaganovich2002,Riemann2005,Riemann2009a,2012b}. The
underlying physics is also the same and is related to the passage of the ion
fluid through the sonic barrier \cite{2012b}. In essence, the two layers
represent the same physical object: the transonic layer, which in one case
assumes the form of a double layer separating the illuminated and
un-illuminated plasmas and in the other case the form of a (positive-charge)
transition layer separating the plasma and a collisionless sheath.

In the form of transition layer separating the plasma and a collisionless
sheath, the transonic layer is difficult to identify in results of numerical
solution of a full problem. In fact, some researchers even believe that the
transition layer is merely an artifact produced by the method of matched
asymptotic expansions. However, the transonic layer at the edge of illuminated
plasma, having the form of a double layer, is easily identifiable; another
proof that the transonic layer is distinguished by specific physical processes
and is therefore not less real than, e.g., the near-wall space-charge sheath.

The normalized ion speed in the un-illuminated plasma in plane geometry given
by the asymptotic analysis equals $1+C_{2}\varepsilon^{1/5}$, where
$C_{2}=0.6714$. Given that $V_{1}\left(  \xi\right)  $ equals $1$ when
extrapolated to $\xi=1$, a natural interpretation is that the ion fluid is
accelerated up to $1$ (the Bohm speed) in the illuminated plasma and from $1$
to $1+C_{2}\varepsilon^{1/5}$ in the double layer. A similar interpretation
applies to potential: the potential difference across the illuminated plasma
equals $-\ln2$ and the potential difference across the double layer is
$-C_{2}\varepsilon^{1/5}$.

The double layers in plane and cylindrical geometries are not qualitatively
different as far as distributions of the electric field and space-charge
densities are concerned. However, distributions of the ion speed are
significantly different: the ion fluid in the double layer is continually
accelerated in cylindrical geometry. This is a consequence of the appearance
in the case $\beta=1$ of an additional term ($\xi_{5}$) on the rhs of
Eq.\ (\ref{12a}), which governs distribution of ion speed in the double layer.
This difference may look somewhat surprising: the double layer is
asymptotically thin, i.e., locally planar, so how can its curvature affect its
structure? Note, however, that this is not the only weak effect affecting the
double layer: the ionization and separation of charges also play a role;
cf.\ the first and second terms on the rhs of Eq.\ (\ref{12a}). The reason for
the latter was discussed in \cite{2012b} and applies to the curvature effect
as well: the balance of forces acting over the ion fluid is delicate in the
vicinity of the sonic point since the main effects (inertia force and the
electrostatic force) cancel; therefore, the weak effects (ionization,
curvature, and separation of charges) also play a role. One should stress in
this connection that Eq.\ (\ref{12a}), describing acceleration of the ion
fluid in the double layer, appears in the second approximation rather than in
the first one.

In plane geometry, the ions enter the near-wall space-charge sheath with a
speed that equals the Bohm speed in the first approximation in $\varepsilon$.
In other words, the Bohm criterion is satisfied with the equality sign, which
is a usual situation, and in the first approximation the near-wall sheath is
the usual Bohm sheath. In other words, the first term of the asymptotic
expansion describing the sheath in the considered problem is exactly the same
as in situations involving an active plasma and a collisionless sheath, for
example, in the problem treated in \cite{Franklin1970}, or in a situation
which would have occurred in the problem considered here if the whole plasma
were illuminated. However, the second terms of the asymptotic expansions
describing the sheath in this problem and in usual situations should be
different, which is due to the different behavior of the second-order term in
the asymptotic zone adjacent to the sheath: in this problem, the second-order
term of expansion of $V$ in the un-illuminated plasma is constant (and equal
to $C_{2}\varepsilon^{1/5}$), while in usual situations the second-order term
of expansion of $V$ in the transition layer has a pole at the wall. Thus, the
near-wall space-charge sheath in the plane partially illuminated plasma is the
usual Bohm sheath to the first approximation in $\varepsilon$ but not to the
second approximation. In cylindrical geometry, the ions enter the sheath with
a speed exceeding the Bohm speed. In other words, the Bohm criterion is
"oversatisfied"; a situation which was known to occur only in a few artificial
models \cite{Riemann1991}.

It has been known for many decades that a (quasi-neutral) presheath, in which
the ion fluid is accelerated up to the Bohm speed, must involve at least one
of the following three mechanisms: ion-atom collisions, ionization, and
multidimensional effects; e.g., \cite{Riemann1991}. As far as ion-atom
collisions and ionization are concerned, this result is somehow
counterintuitive: in terms of ion momentum both ion-atom collisions and
ionization represent a friction force, and why should a friction force be
needed for acceleration?

An explanation of this paradox is as follows. Of course, the ion fluid is
accelerated by the electrostatic force while a friction force has a
decelerating effect; cf.\ Eq.\ (\ref{2a}). On the other hand, the
electrostatic force in a collisionless plane subsonic quasi-neutral plasma
with frozen ionization would exceed the ion inertia force and the ion momentum
balance cannot be ensured; cf.\ Eq.\ (\ref{6}) with the rhs dropped. In other
words, a friction force, while not being the reason of acceleration of a plane
subsonic ion flow under conditions of quasi-neutrality, is its necessary
attribute. The problem of plane partially illuminated plasma offers a
remarkable illustration of this statement: the ion flow is accelerated in the
illuminated plasma, where a friction force due to ionization is present; there
is no friction force in the un-illuminated plasma - and therefore no acceleration.

In cylindrical geometry the rhs of Eq.\ (\ref{8a}) in the illuminated region
equals $-\frac{2}{\xi}V^{2}-\frac{1}{\xi}$ and is negative. In the
un-illuminated plasma the rhs of Eq.\ (\ref{8a}) equals $1/\xi$ and is
positive. In other words, cylindrical geometry provides a retarding effect in
the subsonic region and the ion flow is accelerated due to the presence of the
ionization friction force. The sonic point occurs where the ionization
friction force disappears. Cylindrical geometry provides an accelerating
effect in the supersonic region.

The possibility of experimental testing was discussed in the first part of
this work \cite{2013i}. In this connection, relevant is the case where the
ionization profile is not described by the Heaviside function and decays
smoothly. Asymptotic structure of the solution will remain the same in this
case provided that the decay is fast (occurs on a length scale much smaller
than widths of both illuminated and un-illuminated plasmas). In particular,
there will be a quasi-neutral double layer, where the sonic transition occurs.
The theory of the double layer developed in this work will remain applicable
if length scale of the decay is much smaller than $\varepsilon^{2/5}\Delta$.

\textbf{Acknowledgments} The authors appreciate stimulating discussions with
R. N. Franklin and J. E. Allen. Work at Universidade da Madeira was supported
by FCT - Funda\c{c}\~{a}o para a Ci\^{e}ncia e a Tecnologia of Portugal
through the projects PTDC/FIS-PLA/2708/2012 and PEst-OE/MAT/UI0219/2011.

\appendix{}

\section{ Finding solution for the plane double layer}

\label{Appendix}

In the region $\xi_{5}>0$, Eq.\ (\ref{43}) does not explicitly depend on the
independent variable (is autonomous) and admits an analytical solution, which
may be found as follows. Multiplying this equation by $dV_{5}/d\xi_{5} $ and
integrating, one finds%
\begin{equation}
\frac{V_{5}^{3}}{3}=\frac{1}{2}\left(  \frac{dV_{5}}{d\xi_{5}}\right)
^{2}+C_{2}^{2}V_{5}+C_{4},\label{44}%
\end{equation}
where $C_{4}$ is an integration constant. Solving this equation for $\left(
dV_{5}/d\xi_{5}\right)  ^{-1}$, choosing the sign in the square root with the
use of the assumption that $dV_{5}/d\xi_{5}>0$ and integrating, one obtains
the desired solution in an implicit form
\begin{equation}
\int_{C_{5}}^{V_{5}}\frac{dV_{5}}{\left(  V_{5}^{3}-3C_{2}^{2}V_{5}%
-3C_{4}\right)  ^{1/2}}=\sqrt{\frac{2}{3}}\xi_{5},\label{45}%
\end{equation}
where $C_{5}$ is a new integration constant. Since the solution should exist
for all positive values of $\xi_{5}$, the integral on the lhs of
Eq.\ (\ref{45}) should diverge at a certain value of $V_{5}$. The latter value
should be finite, since the integral converges as $V_{5}\rightarrow\infty$,
and should represents the double root of the cubic polynomial in the
parentheses in the denominator. Such a root exists provided that
$C_{4}=-2C_{2}^{3}/3$ and equals $C_{2}$. Then the integral may be evaluated
analytically and the solution may be transformed to an explicit form:%
\begin{equation}
V_{5}=3C_{2}\tanh^{2}\left(  \sqrt{\frac{C_{2}}{2}}\xi_{5}%
+\operatorname{arctanh}\sqrt{\frac{C_{5}+2C_{2}}{3C_{2}}}\right)
-2C_{2}\text{.}\label{47}%
\end{equation}
The solution involves two integration constants, $C_{2}$ and $C_{5}$, which
have the meaning $C_{2}=V_{5}\left(  \infty\right)  $ and $C_{5}=V_{5}\left(
0\right)  $. It should be the case that $C_{2}>0$, $-2C_{2}<C_{5}<C_{2}$.

Let us now consider Eq.\ (\ref{43}) in the region $\xi_{5}\leq0$. The boundary
condition at $\xi_{5}=0$ is expressed by Eq.\ (\ref{44}) with $C_{4}%
=-2C_{2}^{3}/3$; we recall that not only $V_{5}\left(  \xi_{5}\right)  $ but
also $dV_{5}\left(  \xi_{5}\right)  /d\xi_{5}$ are continuous at $\xi_{5}=0$
as discussed in Sec.\ \ref{The second approximation}. Another boundary
condition is obtained by matching two terms of the expansion (\ref{30}) with
the first term of the expansion (\ref{10}) and reads%
\begin{equation}
\xi_{5}\rightarrow-\infty:\;\;V_{5}=-\sqrt{-2\xi_{5}}+\dots.\label{41}%
\end{equation}
Note that the problem (\ref{43}), (\ref{44}), (\ref{41}) coincides with the
problem describing the intermediate transition layer separating active plasma
and a collisionless sheath \cite{Franklin1970} except for a different boundary
condition at $\xi_{5}=0$. Another substantial difference is in the way in
which Eq.\ (\ref{43}) was derived: in \cite{Franklin1970} it was derived by
means of invoking the next (third) term of the asymptotic expansion
(\ref{30})--(\ref{33}), while in this work it was derived in a straightforward
way from Eq.\ (\ref{ion_mom}); see discussion in \cite{2012b}.

Constant $C_{2}$ may be eliminated from Eq.\ (\ref{43}) by means of
substitution $\xi_{5}=X+C_{2}^{2}/2$:%
\begin{equation}
\frac{d^{2}V_{5}}{dX^{2}}=V_{5}^{2}+2X.\label{47a}%
\end{equation}
Note that this equation coincides with the first Painlev\'{e} equation (e.g.,
\cite{Davis1962,Polyanin2007}) to the accuracy of transformation
$X=3^{1/5}\tilde{X}$, $V_{5}=2\times3^{3/5}Y$.

Asymptotic behavior for $X\rightarrow-\infty$ of solutions of Eq.\ (\ref{47a})
satisfying the matching condition (\ref{41}) may be found to be%
\begin{align}
V_{5} &  =-\left(  -2X\right)  ^{1/2}-\frac{1}{8X^{2}}+\dots+\frac{C_{6}%
}{\left(  -X\right)  ^{1/8}}\exp\left[  i\frac{2^{11/4}}{5}\left(  -X\right)
^{5/4}\right]  \left(  1+\dots\right) \nonumber\\
&  +\frac{C_{7}}{\left(  -X\right)  ^{1/8}}\exp\left[  -i\frac{2^{11/4}}%
{5}\left(  -X\right)  ^{5/4}\right]  \left(  1+\dots\right)  ,\label{16}%
\end{align}
where $C_{6}$ and $C_{7}$ are arbitrary constants. The last two terms on the
rhs of this expression are oscillatory and have to be eliminated in order that
the asymptotic matching be possible beyond the first order; i.e., one should
set $C_{6}=C_{7}=0$. Thus, the proper boundary condition for Eq.\ (\ref{47a})
is given by Eq.\ (\ref{16}) with $C_{6}=C_{7}=0$ and this boundary condition
specifies a unique solution of Eq.\ (\ref{47a}). Note that this conclusion
coincides with the corresponding conclusion in \cite{Franklin1970}, although
Eq.\ (\ref{16}) is not exactly the same.

A convenient way of numerically finding the function $V_{5}$ for $\xi_{5}%
\leq0$ is as follows: Eq.\ (\ref{47a}) is integrated with the initial
condition $V_{5}=-\left(  -2X\right)  ^{1/2}-1/8X^{2}$ from a large negative
$X $ in the direction of increasing $X$ until the condition (\ref{44}) written
in the form%
\begin{equation}
\frac{V_{5}^{3}}{3}=\frac{1}{2}\left(  \frac{dV_{5}}{dX}\right)  ^{2}%
-2XV_{5}-\frac{2}{3}\left(  -2X\right)  ^{3/2}.\label{49}%
\end{equation}
has been met. The value $X=X_{0}$ at which the latter happens corresponds to
$\xi_{5}=0$, so one can find the constants $C_{2}=\sqrt{-2X_{0}}$ and
$C_{5}=V_{5}\left(  X_{0}\right)  $.

This calculation results in the function $V_{5}\left(  X\right)  $ shown in
Fig.\ 1 by the solid line with $X_{0}=-0.2254$, $C_{2}=0.6714$, $C_{5}%
=-0.9262$. Also shown in Fig.\ 1 is the dependence $V_{5}\left(  X\right)  $
in the range $X\geq X_{0}$, described by Eq.\ (\ref{47}); the dash-dotted
line. The dashed line represents data obtained when the numerical calculations
described in the previous paragraph are not stopped when Eq.\ (\ref{49}) has
been satisfied but rather continue until a pole has been encountered; a
procedure identical to the one employed in \cite{Franklin1970}. These data
refer to the transition layer separating active plasma and a collisionless
sheath. The dotted lines represent asymptotic behavior described by the first
two terms on the rhs of Eq.\ (\ref{16}) and by expressions $V_{5}=0.6714$ and
$V_{5}=6\left(  X-X_{1}\right)  ^{-2}$, where $X_{1}$ is the position of the
pole evaluated as $3^{1/5}$ times the value $2.384$ determined numerically in
\cite{Franklin1970}.

\section{Composite asymptotic expansion}

\label{Composite expansions}

The composite expansion of the ion velocity uniformly valid in the illuminated
plasma region to the double layer to the un-illuminated region is obtained by
adding the expansion (\ref{30}) and the first term of the expansion (\ref{10})
and subtracting the common part:%
\begin{equation}
V\left(  \xi,\varepsilon\right)  =\frac{\xi}{1+\sqrt{1-\xi^{2}}}+\left[
1+\varepsilon^{1/5}V_{5}\left(  \xi_{5}\right)  \right]  \ -\left[
1-\sqrt{2\left(  1-\xi\right)  }\right]  ,\label{61}%
\end{equation}
where the first and third terms on the rhs should be discarded for $\xi>1$. In
the illuminated plasma, the second and third terms on the rhs of
Eq.\ (\ref{61}) virtually cancel and the first term is dominating as it
should; in the double layer the first and third terms virtually cancel or are
discarded and the second term is dominating as it should; in the
un-illuminated plasma the first and third terms are discarded and
$V=1+C_{2}\varepsilon^{1/5}$, again as it should be. Error of Eq.\ (\ref{61})
is of the order of $\varepsilon^{2/5}$ from the illuminated plasma to the
double layer to the un-illuminated plasma.

The composite expansion of potential of the same accuracy is obtained in a
similar way and reads%
\begin{equation}
\Phi\left(  \xi,\varepsilon\right)  =\ln\frac{1+\sqrt{1-\xi^{2}}}{2}+\left[
-\ln2-\varepsilon^{1/5}V_{5}\left(  \xi_{5}\right)  \right]  -\left[
-\ln2+\sqrt{2\left(  1-\xi\right)  }\right]  .\label{61a}%
\end{equation}

The first-approximation composite expansion of the electric field is obtained
by differentiating Eq.\ (\ref{61a})\ and reads
\begin{equation}
-\frac{d\Phi}{d\xi}=\frac{\xi}{\sqrt{1-\xi^{2}}+1-\xi^{2}}+\varepsilon
^{-1/5}\frac{dV_{5}\left(  \xi_{5}\right)  }{d\xi_{5}}\ -\frac{1}%
{\sqrt{2\left(  1-\xi\right)  }}.\label{62}%
\end{equation}
Relative error of this formula in the double layer is of the order of
$\varepsilon^{1/5}$.

A deficiency of Eq.\ (\ref{62}) is a discontinuity at $\xi=1$, which stems
from the limit for $\xi\rightarrow1-0$ of the sum of the first and third term
on the rhs being non-zero (it equals $-1$). This discontinuity is of the order
$\varepsilon^{1/5}$ relative to the main term and disappears in the next
approximation. Imagine, for example, that the second term of the expansion of
the electric field in the double layer [i.e., the term associated with the
third term of the expansion (\ref{33})] is $-\frac{1}{2}+\frac{1}{2}%
\tanh\left(  \sqrt{\frac{C_{2}}{2}}\xi_{5}\right)  $. Then the composite
expansion of the electric field would be
\begin{equation}
-\frac{d\Phi}{d\xi}=\frac{\xi}{\sqrt{1-\xi^{2}}+1-\xi^{2}}+\left[
\varepsilon^{-1/5}\frac{dV_{5}\left(  \xi_{5}\right)  }{d\xi_{5}}\ -\frac
{1}{2}+\frac{1}{2}\tanh\left(  \sqrt{\frac{C_{2}}{2}}\xi_{5}\right)  \right]
-\left[  \frac{1}{\sqrt{2\left(  1-\xi\right)  }}-1\right]  .\label{62a}%
\end{equation}
Relative error of this formula in the double layer would be of the order of
$\varepsilon^{2/5}$ and the rhs of Eq.\ (\ref{62a}) is continuous.

In reality, Eq.\ (\ref{62a}) is of course no more accurate than Eq.\ (\ref{62}%
): while the second term of the expansion of the electric field in the double
layer is absent from Eq.\ (\ref{62}), in (\ref{62a}) it is taken into account
in a form which can be only qualitatively correct at best. However,
Eq.\ (\ref{62a}) gives a continuous distribution of the electric field and is
therefore preferable for illustrative purposes.

\begin{center}%
\begin{center}
\includegraphics[
trim=0.000000cm 0.000000cm 0.000000cm 0.094024cm,
height=7.3839cm,
width=8.5172cm
]%
{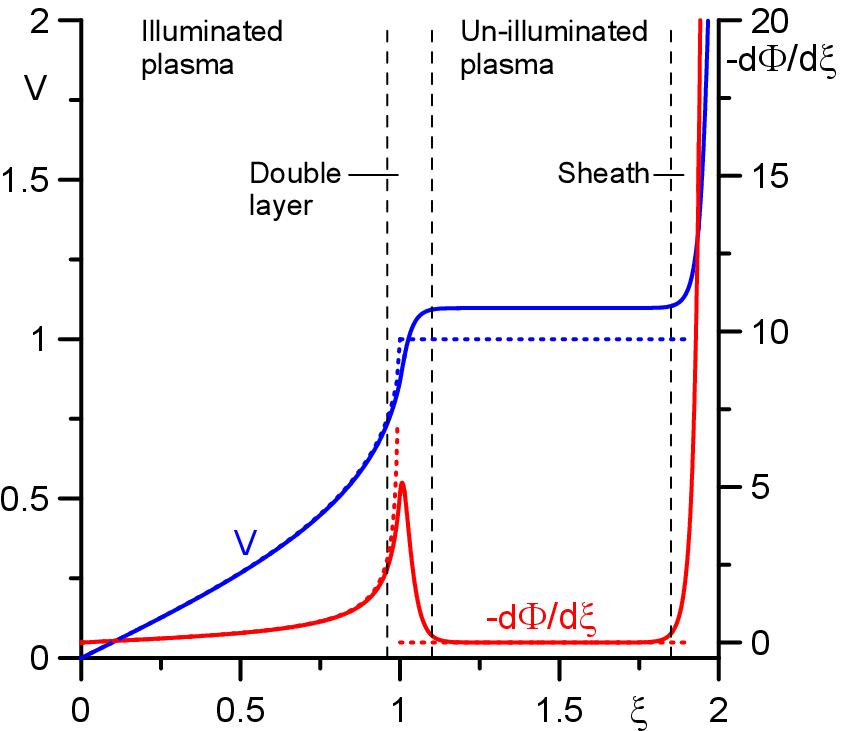}%
\end{center}
\begin{tabular}
[c]{l}%
Fig.\ 1. Distributions of the ion speed and electric field. Solid: numerical\\
calculations, $S=2$, $\varepsilon=0.89\times10^{-4}$. Dotted: patching.
\end{tabular}
\vspace{3cm}%

\begin{center}
\includegraphics[
trim=0.000000cm 0.000000cm 0.000000cm 0.094024cm,
height=7.412cm,
width=7.717cm
]%
{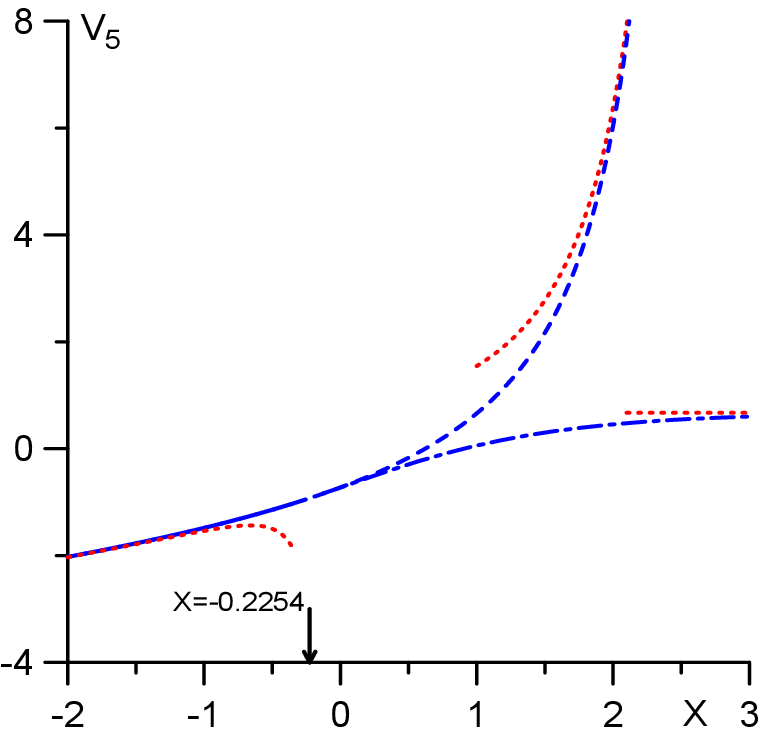}%
\end{center}
\begin{tabular}
[c]{l}%
Fig.\ 2. Asymptotic solution for distribution of ion speed in the vicinity of
the sonic\\
barrier. Plane geometry. Solid + dash-dotted: double layer at the edge of
illuminated\\
plasma. Solid + dashed: transition layer separating active plasma and a
collisionless\\
sheath. Dotted: asymptotic behavior for large $\left\vert X\right\vert $.
\end{tabular}
\newpage%

\begin{tabular}
[c]{cc}%
\raisebox{-0cm}{\includegraphics[
trim=0.000000in 0.000000in 0.002606in 0.000000in,
height=7.4812cm,
width=7.2195cm
]%
{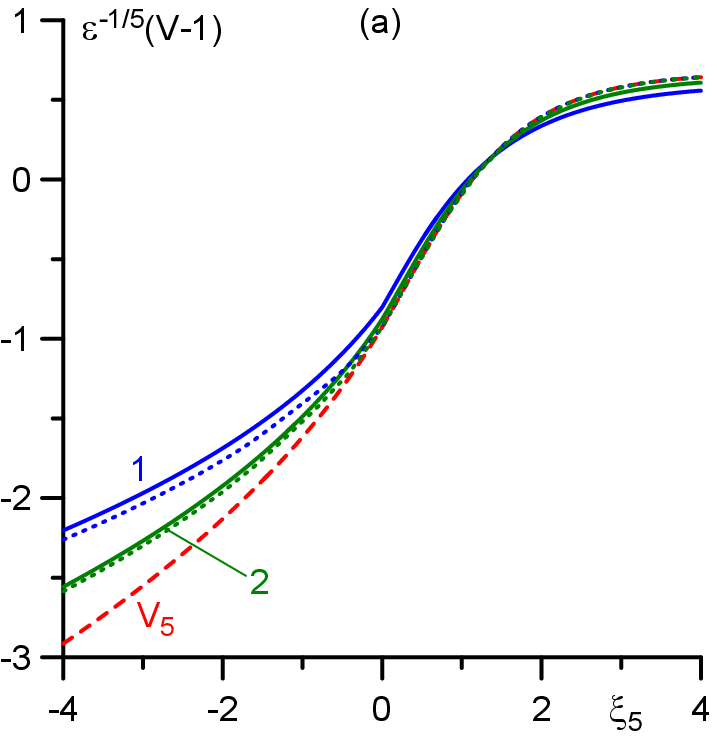}%
}
&
\raisebox{-0cm}{\includegraphics[
trim=0.000000in 0.000000in -0.009772in 0.000000in,
height=7.4812cm,
width=7.2325cm
]%
{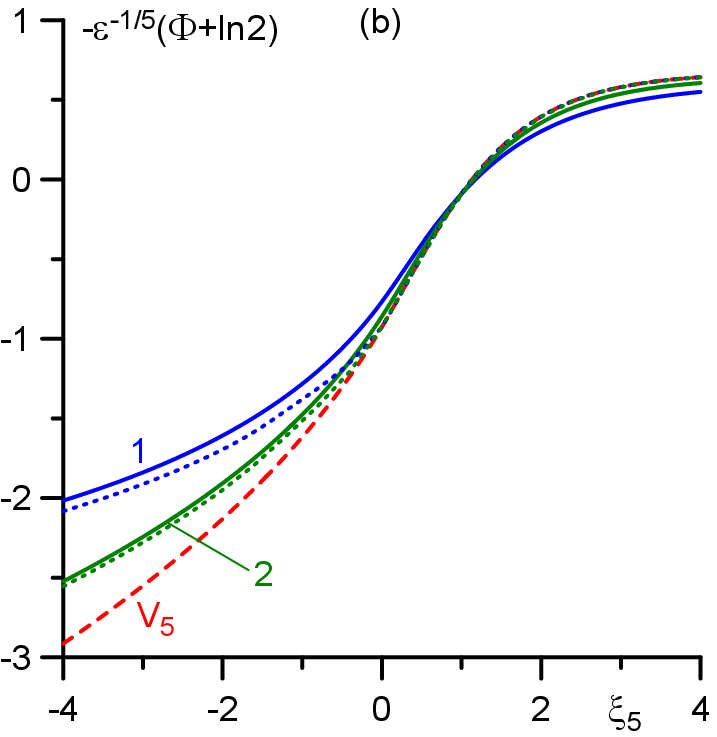}%
}
\\
\medskip & \\%
\raisebox{0.0173cm}{\parbox[b]{7.3969cm}{\begin{center}
\includegraphics[
trim=0.000000in 0.000000in 0.004398in 0.000000in,
height=7.4812cm,
width=7.3969cm
]%
{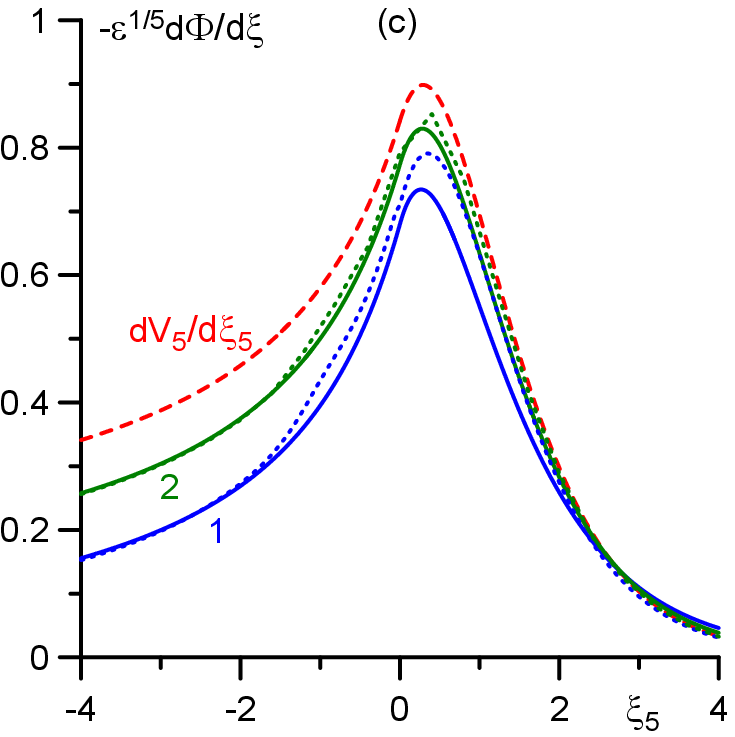}%
\\
{}%
\end{center}}}
&
\raisebox{0.0173cm}{\parbox[b]{7.5266cm}{\begin{center}
\includegraphics[
trim=0.000000in 0.000000in 0.004398in 0.000000in,
height=7.4812cm,
width=7.5266cm
]%
{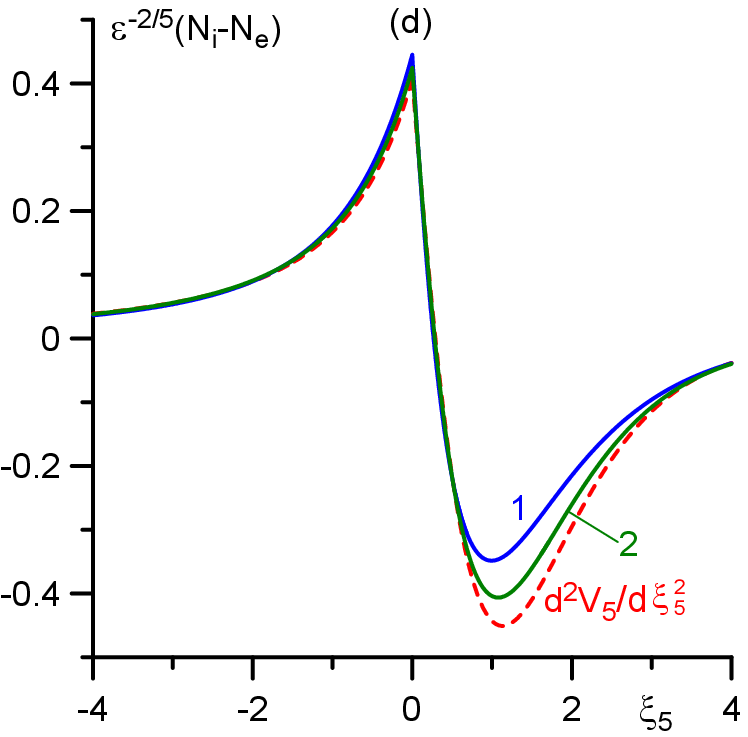}%
\\
{}%
\end{center}}}
\\
\medskip &
\end{tabular}

\begin{tabular}
[c]{l}%
Fig.\ 3. Distributions of normalized parameters in the double layer: the ion
speed (a),\\
the potential (b), the electric field (c), and the density of space charge
(d). Solid:\\
numerical calculations for $S=2$ and $\varepsilon=1.26\times10^{-3}$ (line 1)
or $0.99\times10^{-5}$ (2).\\
Dashed: second-order term of the asymptotic expansion describing the double
layer.\\
Dotted: composite expansion.
\end{tabular}
\newpage

$%
\raisebox{-0cm}{\includegraphics[
trim=0.000000in 0.000000in -0.008770in 0.000000in,
height=7.4812cm,
width=7.4747cm
]%
{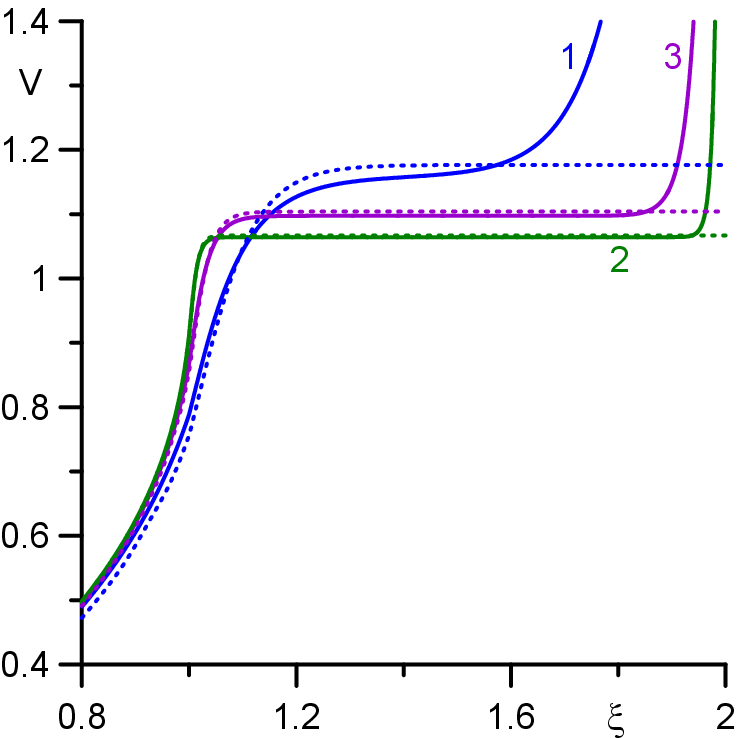}%
}
$%

\begin{tabular}
[c]{l}%
Fig.\ 4. Distributions of ion speed in the double layer and the un-illuminated
plasma.\\
Solid: numerical calculations for $S=2$ and $\varepsilon=1.26\times10^{-3}$
(line 1), $0.99\times10^{-5}$ (2),\\
$0.89\times10^{-4}$ (3). Dotted: composite expansion.
\end{tabular}

\end{center}

\vspace{4cm}

\begin{center}
$%
\raisebox{-0cm}{\includegraphics[
trim=0.000000in 0.000000in -0.008770in 0.000000in,
height=7.4812cm,
width=7.6045cm
]%
{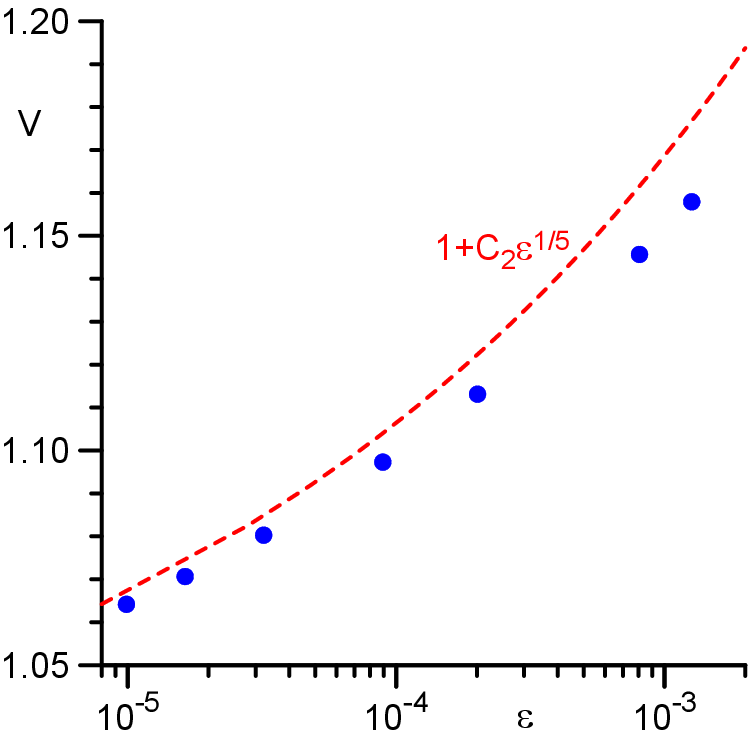}%
}
$%

\begin{tabular}
[c]{l}%
Fig.\ 5. Ion speed in the un-illuminated plasma. Line:\ asymptotic solution.\\
Points: numerical calculations.
\end{tabular}

\end{center}

\end{document}